\documentstyle[12pt,epsfig]{article}

\def\ee{\end{equation}}
\def\be{\begin{equation}}
\def\eea{\end{eqnarray}}
\def\bea{\begin{eqnarray}}
\def\eeas{\end{eqnarray*}}
\def\beas{\begin{eqnarray*}}
 
\begin{document}
\pagenumbering{arabic}
\title{Electron capture on $^{8}B$ nuclei and Superkamiokande results}
\author{F. L. Villante}

\date{}

\maketitle

\small{Dipartimento di Fisica, Universit\`a di Ferrara, I-44100,
Ferrara (Italy)}

\small{Istituto Nazionale di Fisica Nucleare, Sez. di Ferrara,
I-44100, Ferrara (Italy)}

\bigskip

\begin {abstract}

The energy spectrum of recoil electrons from solar neutrino scattering,
as observed by Superkamiokande, is deformed with respect to that
expected from SSM calculations. We considered $\nu-e$ scattering
from neutrinos produced by the electron--capture on $^{8}B$ nuclei,
 $e^{-}+^{8}B\rightarrow^{8}Be^{*}+\nu_{e}$,
as a possible explanation of the spectral deformation. A flux 
$\Phi_{eB}\simeq 10^{4} ~\rm{cm}^{-2}~\rm{s}^{-1}$
could account for Superkamiokande solar neutrino data.
However this explanation is untenable, since the theoretical
prediction, $\Phi_{eB}=(1.3\pm0.2)~\rm{cm}^{-2}~\rm{s}^{-1}$,
is smaller by four orders of magnitude.
\end{abstract}
\smallskip

\bigskip
\bigskip
\bigskip
\bigskip


\pagebreak
The energy spectrum of recoil electrons from solar neutrino scattering,
as reported by Superkamiokande (SK), deviates from Standard Solar Models
(SSM) predictions at energies near and above 
$E_{0}=13 ~\rm{MeV}$
\cite{sk98}.
 This feature can be interpreted in several ways:
\begin{enumerate}
\item[i)] as a distortion of the $^{8}B$ neutrino spectrum,
due to neutrino oscillations between sun and earth \cite{osc1,osc2};
\item[ii)] as an excess of {\it hep} neutrinos\cite{esc,hep}, by about an order
of magnitude with respect to SSM estimates;
\item[iii)] as a combination of the two solutions above.
\end{enumerate}
The problem of neutrino oscillations
is so important that any alternative explanation of the data,
although unlikely, has to be investigated carefully.
In this spirit  we consider the case of neutrinos from
electron-capture on $^{8}B$ nuclei:
\be
e^{-}+^{8}B\rightarrow^{8}Be^{*}+\nu_{e}\rightarrow2\alpha+\nu_{e} ~,
\label{reac}
\ee
as a possible source of the spectral distortion observed by SK.
The energy spectrum of these neutrinos, which we refer to as {\it eB}
neutrinos, is peaked near $E_{eB}=15.5~\rm{MeV}$ \cite{spectrum2} with a
full width half maximum $\Delta=1.4~\rm{MeV}$, see fig.1.
The energy spectrum of recoil electrons from {\it eB} neutrinos
is pratically flat up to about $E_{eB}-m_{e}/2$, contrary to
that from $^{8}B$ neutrino scattering, which is a decreasing function
of electron energy and vanishes near $14 ~\rm{MeV}$.
 A substantial flux of {\it eB} neutrinos could then mimic the
shape of the electron spectrum reported by SK.

 In section 1, we look quantitatively
at this idea, determining how many 
{\it eB} neutrinos are required to account for SK data. In section
2, we compare the result obtained with the theoretical predictions
for the {\it eB} neutrino flux.

\section{How many {\it eB} neutrinos are needed?}

SK has recently presented a measurement
of the energy spectrum of recoil electrons from solar
neutrino scattering, corresponding to 504 days of data taking \cite{sk98}.
 By assuming the SSM estimate of the {\it hep} neutrino flux \cite{bp98},
$\Phi_{hep}^{SSM}\simeq2\times10^{3}~\rm{cm}^{-2}~\rm{s}^{-1}$ and an undeformed
$^{8}B$ neutrino spectrum, with an arbitrary normalization,
they obtained a $\chi^2/D.O.F.=25.3/15$, corresponding to
a 4.6 \% confidence level \cite{sk98}. The poor fit 
is due mainly to the behaviour of the energy-bins above $13~\rm{MeV}$.

Escribano et al. \cite{esc} suggested that a {\it hep} flux significantly
larger than the SSM estimate could reproduce the observed spectrum.
Bahcall et al. \cite{hep} have shown that a flux $\Phi_{hep}
\ge 20 \times \Phi_{hep}^{SSM}$ could
actually mimic the SK spectrum.

 Alternatively, one can keep
the SSM prediction for {\it hep} neutrinos and look for other
high energy neutrino sources. Since the average energy of {\it eB} neutrinos
is roughly twice than that of {\it hep} neutrinos and since the
$\nu-e$ scattering cross section
increase linearly with energy, one expects that a flux 
$\Phi_{eB}\simeq 10\times\Phi_{hep}^{SSM}\simeq 2\times 10^{4}~\rm{cm}^{-2}
~\rm{s}^{-1}$ could be sufficient to account for the high energy behaviour
of SK data.

In order to provide a quantitative estimate, let us analyse the data
by using as free parameters $\alpha=\Phi_{B}/\Phi_{B}^{SSM}$ and
$\delta=\Phi_{eB}/\Phi_{B}^{SSM} $, where $\Phi_{B}^{SSM}=5.15
\times10^{6}~\rm{cm}^{-2}~\rm{s}^{-1}$ is the SSM prediction for
the $^{8}B$ neutrino flux \cite{bp98}. We define, 
in analogy with \cite{sk98}, the following $\chi^{2}$:
\be
\chi^{2}=\sum_{i=1}^{16}
\left\{\frac{\frac{R_{i}}{SSM_{i}}
-\frac{\alpha+\delta\times B_{i}/SSM_{i}}
{(1+\delta_{i,exp}\times\beta)(1+\delta_{i,cal}\times\gamma)}}
{\sigma_{i}} \right\} ^{2}
+\gamma^{2}+\beta^{2}\ .
\label{chi2}
\ee
In the previous relation $R_{i}$ is the number of
solar neutrino events observed in the i-th energy-bin;
$SSM_{i}$ \footnote{The quantities $SSM_{i}$ and $B_{i}$ have been calculated
taking into account the energy resolution of SK \cite{skres},
as described e.g. in \cite{noi}}
 is the number of events
in the same energy bin due to $^{8}B$ neutrinos, for a total
flux $\Phi_{B}^{SSM}$;
$B_{i}$ is the same number due to {\it eB} neutrinos, again for a total
flux $\Phi_{B}^{SSM}$;
the quantities $\delta_{i,exp}$, $\delta_{i,cal}$, $\sigma_{i}$,
defined as in \cite{sk98}, take into account correlated
and uncorrelated theoretical and experimental errors;
 the free parameters $\beta$ and $\gamma$ are used for
constraining the variation of correlated systematic errors.
For each value of $\delta$ we determined the parameters
$\alpha$, $\beta$ and $\gamma$ so as to determine the minimum of
eq. (\ref{chi2}), $\chi^{2}_{m}$, see fig. 2.

As expected, a large {\it eB} neutrino flux produces a steep increase
in the high energy tail of the Superkamiokande normalized spectrum, see fig.3.
The best-fit is obtained when $\Phi_{eB}=1.1\times10^{4} 
\rm{cm}^{-2}\rm{s}^{-1}$,
corresponding to $\chi^{2}_{min}/D.O.F.=15.7/14$.
Acceptable fits are anyhow obtained for $\Phi_{eB}$ in the range
$(0.3-2)\times 10^{4} \rm{cm}^{-2}\rm{s}^{-1}$, see fig. 2.

\section{Theoretical evaluation of the {\it eB} neutrino flux}

Boron ($^8$B) is produced in the sun,
 according to the following reaction 
$$ ^7 Be + p \rightarrow ^8B + \gamma$$
and it undergoes $\beta^+$ decay
\be
^8B \rightarrow ^8 Be^* + e^+ +\nu_e \rightarrow 2\alpha + e^+ +\nu_e
\label{decay}
\ee
or electron capture reaction
\be
^8B + e^-\rightarrow ^8 Be^* + \nu_e \rightarrow 2\alpha + \nu_e ~.
\label{EC}
\ee
The process under consideration is an allowed transition: in fact (see 
ref.\cite{Led78}) $ J^P (^8B) = J^P (^8Be^*) = 2^+$.
In this case, the ratio $R$ between electron capture probability
($\Gamma_{eB}$) 
and $\beta^+ $ decay probability ($\Gamma_{\beta^+}$)
does not depend on the matrix elements of the transition operator between the 
nuclear states.
A simple phase--space calculation, assuming that the electron
number density at nuclear site $n_{e}(0)$ can be approximated by
the average electron number density $n_{e}$, gives immediately
\be
R=
\frac{1}{8 \pi}\left(\frac{hc}{m_e c^{2}}\right)^{3}
\times\left(\frac{E_{eB}}{m_{e}c^{2}}\right)^{2}\times
f^{-1}\times{n}_{e} ~,
\label{ratio}
\ee
where, for later convenience, we show
explicity the dimensionless phase--space factor
associated to $\beta^{+}$ decay,
$f\simeq[(E_{eB}-m_{e}c^{2})/m_{e}c^{2}]^{5}/30\simeq7.1\times10^{5}$.
For $n_{e}\simeq5.4\times10^{25}\rm{cm}^{-3}$ as suggested by SSM,
one has $R\simeq4\times10^{-8}$ and consequently
\be
\Phi_{eB}=R\times\Phi_{B}^{SSM}=2\times10^{-1}~\rm{cm}^{-2}~\rm{s}^{-1} ~,
\ee
i.e. five orders of magnitude lower than that required to account for SK data.

It is anyhow useful to estimate $\Phi_{eB}$ with a better accuracy.
With respect to the naive estimate given previously, one
should consider the effects of interactions with the solar plasma. 
The distortion of the positron wave function in the
$\beta^{+}$ decay rate can be described as a modification of
the dimensionless phase--space factor $f$, which is now given by
$f=5.70\times 10^{5}$ \cite{ec0,ec1}.
Moreover
the electron density at nucleus $n_{e}(0)$ is larger than $n_{e}$
and, consequently, the ratio $R$ has to be enhanced, with
respect to eq. (\ref{ratio}), by a factor
\be
\omega=\frac{n_{e}(0)}{n_{e}}
\label{omega}
\ee
For a precise estimate of $\omega$
one has to take into account:
{\it i)} distortion of electron wave functions in the Coulomb
field of nucleus\cite{bah62},
{\it ii)} electron capture from bound states \cite{iks},
{\it iii)} screening effects \cite{iks,bm}.
Let us discuss the problem in some detail, following the
lines of Gruzinov and Bahcall who recently produced a clear and comprehensive
analysis of the $^{7}Be$ electron capture in the sun \cite{gb}:
\begin{enumerate}
\item[\it i)]
Because of the Couloumb field of the nucleus, the wave functions of
continuum electron states differ from plane waves. The rate of electron capture
from continuum has then to be corrected by an enhancement
factor $\omega_{c}$ \cite{bah62}:
\be
\omega_{c} = <\left|\frac{\psi_{coul}(0)}{\psi_{free}(0)}\right|^{2}>=
\left(\frac{m_{e}c^2}{kT}\right)^{\frac{1}{2}}\times
\left(Z\alpha \right)\times 2 \times \left(2\pi\right)^{\frac{1}{2}} 
\times I(\beta) ~,
\ee
where the average is taken over electron thermal distribution.
In the previous relation $T$ is the Sun temperature, while $I(\beta)$
is a correction factor of order unity, defined e.g. in \cite{ec2}.
For $R/R_{\odot}\simeq0.05$, which corresponds to the solar region where
the production of $^{8}B$ neutrinos is maximal,
the density enhancement at nucleus due to electron in continuum states is
$\omega_{c}=3.82$.
\item[\it ii)]
As pointed out by Iben, Kalata e Schwartz \cite{iks},
under solar conditions bound electrons give 
a substantial contribution to the electron density at the nucleus.
The bound state enhancement factor is given by \cite{gb}:
\be
\omega_{b} = \pi^{\frac{1}{2}}\times\hbar^{3}\times
\left(\frac{m_{e}kT}{2}\right)^{-\frac{3}{2}}
\sum_{n} \left(\frac{Z}{a_{0}n}\right)^{3}
\exp\left(Z^{2}e^{2}/2n^{2}a_{0}kT\right) ~,
\ee
where $a_{0}$ is the Bohr radius. For $R/R_{\odot}\simeq0.05$, the
bound state enhancement factor is $\omega_{b}=2.94$. The total density
enhancement factor is then $\omega_{c}+\omega_{b}=6.76$
\item[\it iii)]
Screening effects reduce the electron density at
nucleus for both bound \cite{iks} and continuum electron states \cite{bm}.
 If the temperature is
sufficiently high and if the screened potential
can by described by
\be
V(r)=-\frac{Ze^{2}}{r}\exp(-r/R_{D}) ~,
\label{salpeter}
\ee
where $R_{D}$ is the Debye radius,
by using a thermodynamical argument one finds \cite{gb}:
\be
\omega=\exp(-Ze^{2}/kTR_{D})\times(\omega_{b}+\omega_{c}) ~.
\label{screen}
\ee
For $R/R_{\odot}\simeq0.05$ the total density enhancement factor
, due to screening effects, is reduced to $\omega=5.34$.
The small difference between this value of $\omega$
and that given by \cite{gb} is due to the fact that they
have been calculated for slightly different solar regions.
Relation (\ref{screen}) is not so straightforward, especially
because of the possible inadequacies
of the Debye screening theory \cite{joh} and because
of the relatively large thermal fluctuations which could results
from the small number of ions in a Debye sphere \cite{sal}.
For the similar case of $^{7}Be$ electron capture, 
Gruzinov \& Bahcall have performed a detailed analisys
of the problem, concluding that relation (\ref{omega}) is
accurate at the 2\% level. 
\end{enumerate}

By using the previous relations we can determine the ratio between
electron capture and $\beta^{+}$ decay rates.
We obtain:
\be
R=2.6\times 10^{-7}\times(1\pm0.02)
\ee
This value is about 30\% larger than previous estimates \cite{ec1}
which took into account only continuum electron states contribution.
By using the SSM estimate of the $^{8}B$ neutrino flux,
which is uncertain by about 17\% \cite{bp98}, one concludes
\be
\Phi_{eB}=R\times\Phi_{B}^{SSM}=1.3\times(1\pm0.17)~\rm{cm}^{-2}\rm{s}^{-1} ~.
\ee
The predicted neutrino flux is
lower by a factor $10^{4}$  than required
to account for SK data and the calculation method is robust.
We conclude that {\it eB}
neutrinos cannot explain the spectral distributions of solar
neutrino events reported by SK.

 The author thanks
 M.R. Quaglia, G. Fiorentini, P. Pizzochero and P. Bortignon for
useful discussions and for earlier
collaboration on the subject of this paper.

\begin{figure}[htb]
\epsfig{file=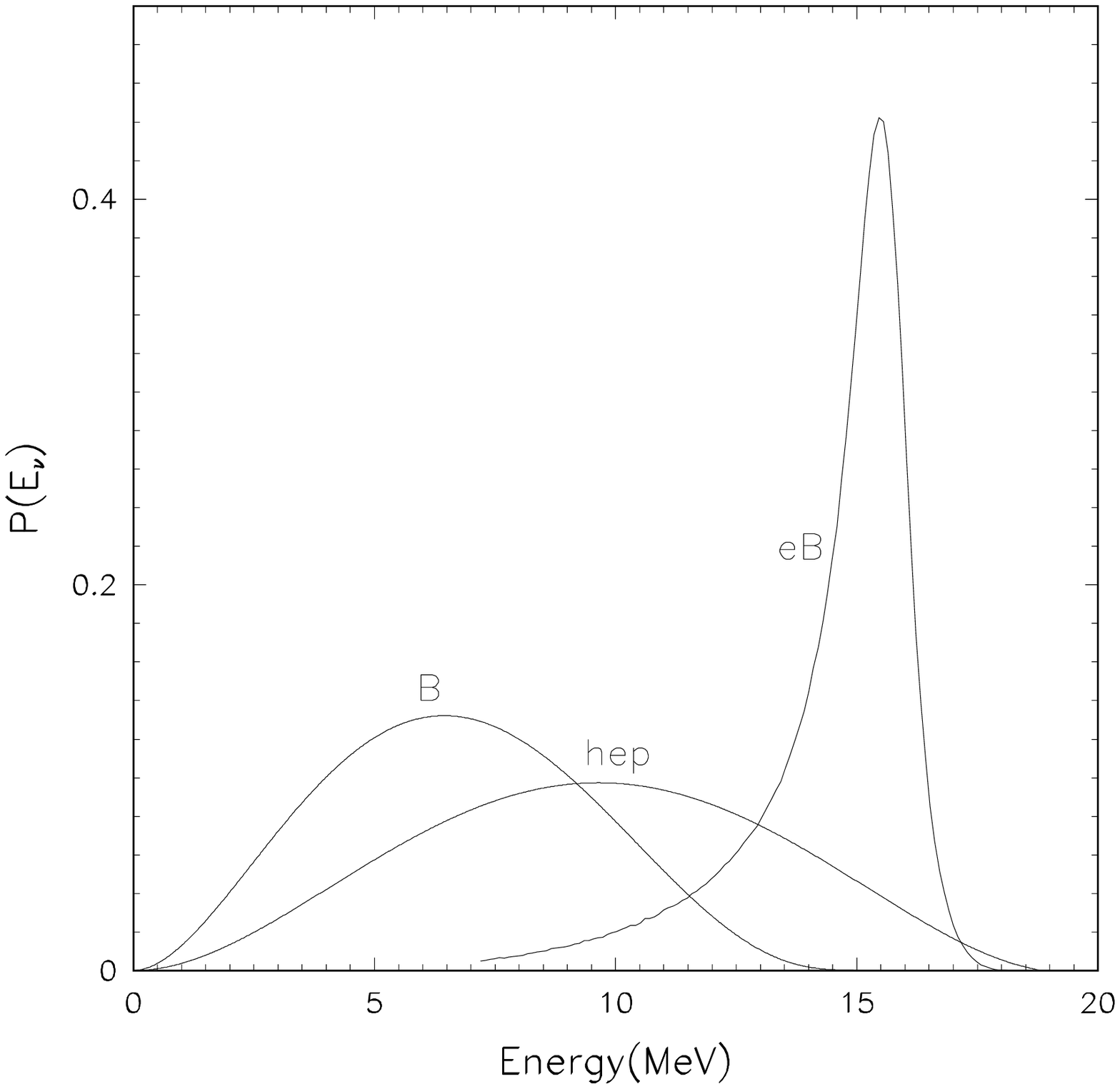,height=20cm,width=15cm}
\caption{Normalized energy spectra of $^{8}B$, hep and eB neutrinos.}
\label{fig1}
\end{figure}

\begin{figure}[htb]
\epsfig{file=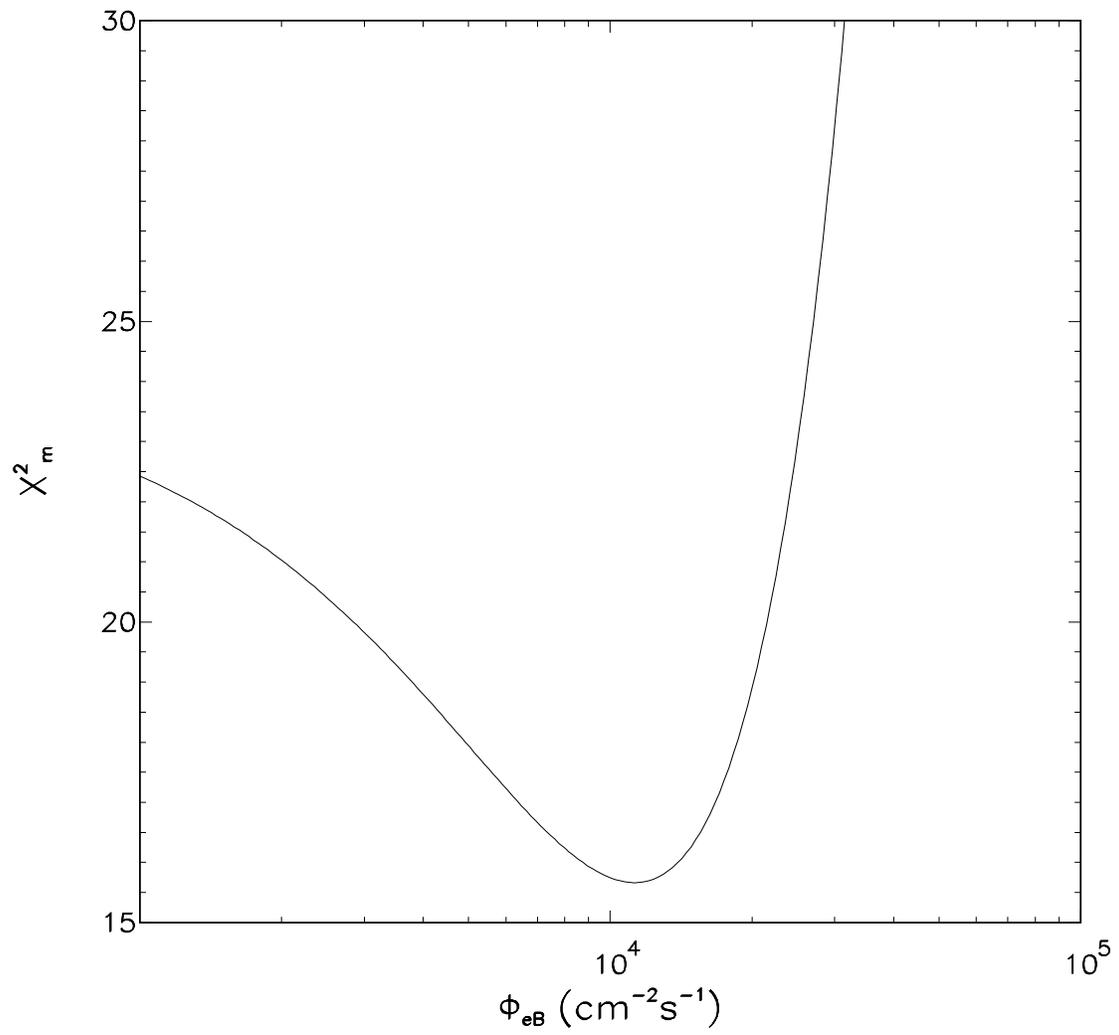,height=20cm,width=15cm}
\caption{The minimum of eq.(\ref{chi2}) for a given value of $\Phi_{eB}$.}
\label{fig2}
\end{figure}

\begin{figure}[htb]
\epsfig{file=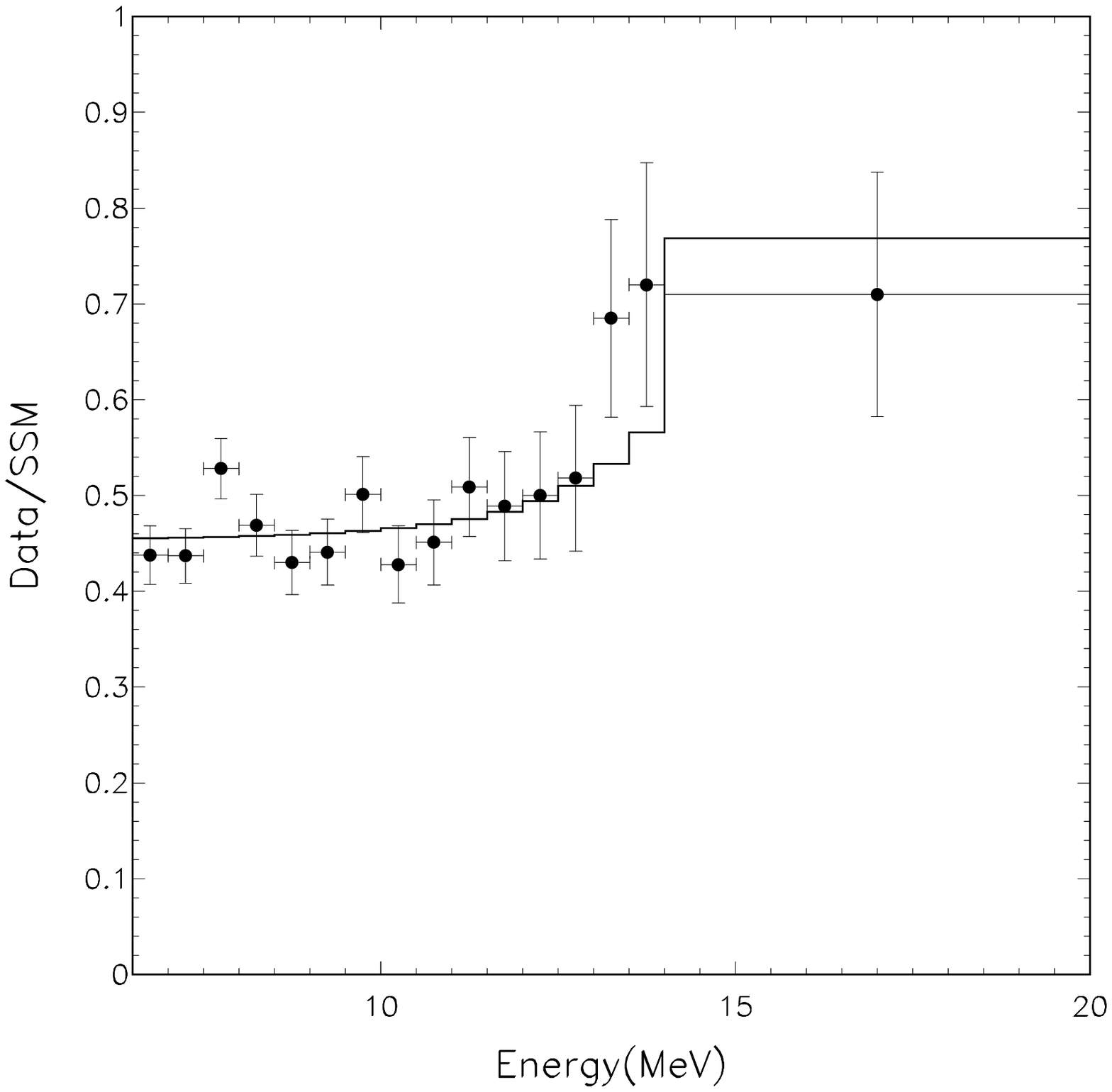,height=20cm,width=15cm}
\caption{Observed electron energy spectrum normalized to
SSM expectations (dots). The solid line is the prediction for
$\Phi_{eB}=1.1\times10^{4}~\rm{cm}^{-2}\rm{s}^{-1}$.}
\label{fig3}
\end{figure}


\begin{thebibliography}{99}

\bibitem{sk98}
Super-Kamiokande Collaboration (Y. Fukuda et al.), ICRR-REPORT-442-98-38,
Dec 1998, hep-ex/9812011



\bibitem{osc1}
J.N. Bahcall, P. I. Krastev, A. Yu. Smirnov, Phys. Rev D 58 (1998) 096016

\bibitem{osc2}
V. Berezinsky, G. Fiorentini, M. Lissia, hep-ph/9811352

\bibitem{esc}
R. Escribano, J. M. Frere, A. Gevaert, D. Monderen, 
Phys. Lett. B 444 (1998) 397

\bibitem{hep}
J.N. Bahcall, P. I. Krastev, Phys. Lett. B, 436 (1998) 243

\bibitem{bp98}
J.N. Bahcall, S. Basu, M. H. Pinsonneault, Phys. Lett. B 433 (1998) 1

\bibitem{spectrum2}
E. K. Warburton, Phys. Rev. C 33 (1986) 303

\bibitem{skres}
Y. Suzuki, in {\em Neutrino~'98}, 
XVIII International Conference on Neutrino Physics
and Astrophysics, Takayama, Japan, June 1998, to
appear in the Proceedings. Scanned transparencies
available at the URL 
http://www-sk.icrr.u-tokyo.ac.jp/nu98/scan/index.html
~.

\bibitem{noi}
F. L. Villante, G. Fiorentini, E. Lisi, Phys. Rev. D 59 (1999) 013006

\bibitem{Led78}
C. M. Lederer and V. S. Shirley.
\newblock {\em Tables of Isotopes}.
\newblock John Wiley and Sons, Inc., 1978.

\bibitem{Bam77}
W.~Bambynek {\it et al.}
\newblock {\em Rev. Mod. Phys.}, {\bf 49}, 77, 1977.

\bibitem{ec0}
J. N. Bahcall, Nuclear Physics  75 (1966) 10

\bibitem{ec1}
J. N. Bahcall, Phys. Rev. D 41 (1990) 2964\\
J. N. Bahcall, Phys. Rev. 135 (1964) B137

\bibitem{ec2}
J. N. Bahcall, R. M. May, ApJ 155 (1969) 501

\bibitem{bah62}
J. N. Bahcall, Phys. Rev. 128 (1962) 1297

\bibitem{iks}
I. Iben, K. Kalata, J. Schwartz, ApJ 150 (1967) 1001

\bibitem{bm}
J. N. Bahcall, C. P. Moeller, ApJ 155 (1969) 511

\bibitem{gb}
A. V. Gruzinov, J. N. Bahcall ApJ 490 (1997) 437

\bibitem{joh}
C. W. Johnson, E. Kolbe, S. E. Koonin, K. Langanke, ApJ (1992) 320

\bibitem{sal}
W. D. Watson, E. E. Salpeter, ApJ 181 (1973) 237



\end{thebibliography}
\end{document}